\documentclass[11pt,dvips]{article}
\usepackage{psfig} 
\makeatletter
\renewcommand{\section}{\@startsection{section}{1}{0in}
	{0.4\baselineskip}{0.1\baselineskip}{\Large\bf}}
\renewcommand{\subsection}{\@startsection{subsection}{2}{0in}
	{0.25\baselineskip}{-\baselineskip}{\large\bf}}
\renewcommand{\subsubsection}{\@startsection{subsubsection}{3}{0in}
	{0.1\baselineskip}{-\baselineskip}{\normalsize\bf}}

\makeatother
\begin{document}

%
\thispagestyle{myheadings}
%
\markright{OG 2.2.09}
%
%
\newcount\eqnno
\eqnno=0
\def\eqnum{\global\advance \eqnno by 1 \eqno{(\the\eqnno)}}
\def\eqnumal{\global\advance \eqnno by 1   (A\the\eqnno)}
\def\labeleqn#1{\global\advance \eqnno by 1\xdef#1{\the\eqnno }
    \global\advance \eqnno by -1}
{\catcode`\@=11                                                  
\gdef\SchlangeUnter#1#2{\lower2pt\vbox{\baselineskip 0pt\lineskip0pt    
\ialign{$\m@th#1\hfil##\hfil$\crcr#2\crcr\sim\crcr}}}}           
\def\gtrsim{\mathrel{\mathpalette\SchlangeUnter>}}               
\def\lesssim{\mathrel{\mathpalette\SchlangeUnter<}}    
%
%
%
\def\frac#1#2{{ #1 \over { #2 }}}
\def\tacc{t_{\rm acc}}
\def\tSNR{t_{\rm SNR}}
\def\EoverP{(e/p)_{\rm 10GeV}}
\def\EnSN{{\cal E}_{\rm SN}}
\def\itt{\it}
\def\bff{\bf}
\def\aa#1#2#3{ 19#1, {\itt A\&A,} {\bff #2}, #3} 
\def\aapress{{\itt Ap. J.,} in press}
\def\aas#1#2#3{ 19#1, {\itt A\&AS,} {\bff #2}, #3} 
\def\aasubmit#1{ 19#1, {\itt A\&A,} submitted.}
\def\aasup#1#2#3{ 19#1, {\itt A.A. Suppl.,} {\bff #2}, #3} 
\def\agumonofour#1{ 1985, in {\itt Collisionless Shocks in the
   Heliosphere: A Tutorial Review,} AGU Vol. 34, 
   p. #1, ed. by R.G. Stone and B.T. Tsurutani,  Washington,
   D.C.}
\def\aj#1#2#3{ 19#1, {\itt A.J.,} {\bff #2}, #3}
\def\anngeophys#1#2#3{ 19#1, {\itt Ann. Geophys.,} {\bff #2}, #3} 
\def\annrev#1#2#3{ 19#1, {\itt Ann. Rev. Astr. Ap.,} {\bff #2}, #3}
\def\apj#1#2#3{ 19#1, {\itt Ap.J.,} {\bff #2}, #3} 
\def\apjlet#1#2#3{ 19#1, {\itt Ap.J.(Letts),} {\bff  #2}, #3} 
\def\apjpress{{\itt Ap. J.,} in press}
\def\apjsubmit{{\itt Ap. J.,} submitted}
\def\apjs#1#2#3{ 19#1, {\itt Ap.J.Suppl.,} {\bff #2}, #3} 
\def\app#1#2#3{ 19#1, {\itt Astroparticle Phys.,} {\bff #2}, #3} 
\def\asr#1#2#3{ 19#1, {\itt Adv. Space Res.,} {\bff #2}, #3}
\def\araa#1#2#3{ 19#1, {\itt Ann. Rev. Astr. Astrophys.,} {\bff #2}, 
   #3}
\def\ass#1#2#3{ 19#1, {\itt Astr. Sp. Sci.,} {\bff #2}, #3}
\def\eos#1#2#3{ 19#1, {\itt EOS,} {\bff #2}, #3}
\def\icrcplovdiv#1#2{ 1977, in {\itt Proc. 15th ICRC(Plovdiv)}, 
   {\bff #1}, #2} 
\def\icrcparis#1#2{ 1981, in {\itt Proc. 17th ICRC(Paris)}, 
   {\bff #1}, #2} 
\def\icrcbang#1#2{ 1983, in {\itt Proc. 18th ICRC(Bangalore)}, 
   {\bff #1}, #2} 
\def\icrclajolla#1#2{ 1985, in {\itt Proc. 19th ICRC(La Jolla)}, 
   {\bff #1}, #2} 
\def\icrcmoscow#1#2{ 1987, in {\itt Proc. 20th ICRC(Moscow)}, 
   {\bff #1}, #2} 
\def\icrcadel#1#2{ 1990, in {\itt Proc. 21st ICRC(Adelaide)}, 
   {\bff #1}, #2} 
\def\icrcdub#1#2{ 1991, in {\itt Proc. 22nd ICRC(Dublin)}, 
  {\bff #1}, #2} 
\def\icrccalgary#1#2{ 1993, in {\itt Proc. 23rd ICRC(Calgary)}, 
  {\bff #1}, #2} 
\def\icrcrome#1#2{ 1995, in {\itt Proc. 24th ICRC(Rome)}, 
  {\bff #1}, #2} 
\def\icrcromepress{ 1995, {\itt Proc. 24th ICRC(Rome)}, in press}
\def\grl#1#2#3{ 19#1, {\itt G.R.L., } {\bff #2}, #3}
\def\jcp#1#2#3{ 19#1, {\itt J. Comput. Phys., } {\bff #2}, #3}
\def\JETP#1#2#3{ 19#1, {\itt JETP, } {\bff #2}, #3}
\def\JETPlet#1#2#3{ 19#1, {\itt JETP Lett., } {\bff #2}, #3}
\def\jgr#1#2#3{ 19#1, {\itt J.G.R., } {\bff #2}, #3}
\def\jpG#1#2#3{ 19#1, {\itt J.Phys.G: Nucl. Part. Phys., } 
     {\bff #2}, #3}
\def\mnras#1#2#3{ 19#1, {\itt M.N.R.A.S.,} {\bff #2}, #3}
\def\nature#1#2#3{ 19#1, {\itt Nature,} {\bff #2}, #3} 
\def\nucphysB#1#2#3{ 19#1, {\itt Nucl. Phys. B (Proc. Suppl.,} 
    {\bff #2}, #3} 
\def\pss#1#2#3{ 19#1, {\itt Planet. Sp. Sci.,} {\bff #2}, #3}
\def\pf#1#2#3{ 19#1, {\itt Phys. Fluids,} {\bff #2}, #3}
\def\phyrepts#1#2#3{ 19#1, {\itt Phys. Repts.,} {\bff #2}, #3}
\def\pr#1#2#3{ 19#1, {\itt Phys. Rev.,} {\bff #2}, #3}
\def\prD#1#2#3{ 19#1, {\itt Phys. Rev. D,} {\bff #2}, #3}
\def\prl#1#2#3{ 19#1, {\itt Phys. Rev. Letts,} {\bff #2}, #3}
\def\pasj#1#2#3{ 19#1, {\itt Pub. Astro. Soc. Japan,} {\bff #2}, #3}
\def\pasp#1#2#3{ 19#1, {\itt Pub. Astro. Soc. Pac.,} {\bff #2}, #3}
\def\rpp#1#2#3{ 19#1, {\itt Rep. Prog. Phys.,} {\bff #2}, #3}
\def\revgeospphy#1#2#3{ 19#1, {\itt Rev. Geophys and Sp. Phys.,} 
   {\bff #2}, #3}
\def\rmp#1#2#3{ 19#1, {\itt Rev. Mod. Phys.,} {\bff #2}, #3}
\def\science#1#2#3{ 19#1, {\itt Science,} {\bff #2}, #3}
\def\sp#1#2#3{ 19#1, {\itt Solar Phys.,} {\bff #2}, #3}
\def\ssr#1#2#3{ 19#1, {\itt Space Sci. Rev.,} {\bff #2}, #3}
\def\reff{\hbox{}\par\noindent \hangafter=1 \hangindent=5.0truemm}
\def\EB{Ellison \& Berezhko }
\def\EBJ{Ellison, Baring, \& Jones }
\def\EJR{Ellison, Jones, \& Reynolds }
\def\JE{Jones \& Ellison }
\def\ALS{Axford, Leer, \& Skadron }
\def\BE{Berezhko \& Ellison }
\def\BO{Blandford \& Ostriker }
\def\EE{Ellison \& Eichler }
\def\EMP{Ellison, M\"obius, \& Paschmann }
%
%
\def\Ek{E_k}
\def\dsa{diffusive shock acceleration}
\def\syn{synchrotron}
\def\IC{inverse Compton}
\def\ic{inverse Compton}
\def\brem{bremsstrahlung}
\def\brems{bremsstrahlung}
\def\pion{pion-decay}
\def\SSC{\syn-self-Compton}
\def\mc{Monte Carlo }
\def\mpc{m c}
\def\etaMC{\eta_{\rm mc}}
\def\rg{r_g}
\def\Pcal{\cal P}
\def\PgasZ{P_{g0}}
\def\PgasOne{P_{g1}}
\def\Eff{\epsilon(>E)}
\def\EffRel{\epsilon_{\rm rel}}
\def\mp{m}
\def\Mco{M_{c0}}
\def\CR{cosmic ray }
\def\gamg{\gamma_g}
\def\gamgas{\gamma_g}
\def\rhooz{ {\rho_1 \over {\rho_0}}}
\def\rhozone{ {\rho_0 \over {\rho_1}}}
\def\alf{Alfv\'en }
\def\mcchi{\chi mc}
\def\mcochi{mc/ \chi}
\def\FEnP{F^{\prime}_{\rm E}}
\def\FEnPP{F^{\prime\prime}_{\rm E}}
\def\RelFrac{F^{\prime}_{\rm rel}}
\def\distrFunc{f({\bf p})}
\def\eg{e.g. }
\def\egc{e.g., }
\def\ie{i.e. }
\def\iec{i.e., }
\def\etal{et al.~}
\def\etalc{et al., }
\def\cl#1{\centerline{#1}}
\def\alfvel{c_A}
\def\Msun{M_{\odot}}
\def\Mej{M_{\rm ej}}
\def\Vsk{V_{\rm sk}}
\def\Rsk{R_{\rm sk}}
\def\Rtot{r_{\rm tot}}
\def\rtot{r_{\rm tot}}
\def\Rtotmax{r_{\rm tot}^{\rm max}}
\def\rtotA{r_{\rm tot}^A}
\def\RtotF{r_{\rm tot}(\Fesc)}
\def\rsub{r_{\rm sub}}
\def\rsubA{r_{\rm sub}^A}
\def\Rsub{r_{\rm sub}}
\def\Rtwo{r_{\rm 2}}
\def\qsub{q_{\rm sub}}
\def\qint{q_{\rm int}}
\def\qmin{q_{\rm min}}
\def\qTP{q_{\rm tp}}
\def\qtp{q_{\rm tp}}
\def\kmps{ \hbox{km~s$^{-1}$}}
\def\pcc{ \hbox{cm$^{-3}$}}
\def\deg{^\circ}
\def\muG{\mu{\rm G}}
\def\rg{r_{\rm g}}
\def\denPro{n_{\rm H}}
\def\denProZ{n_{\rm H,0}}
\def\denProOne{n_{\rm H,1}}
\def\denHe{n_{\rm He}}
\def\denHeOne{n_{\rm He,1}}
\def\denElecOne{n_{\rm e,1}}
\def\TempHZ{T_{\rm H,0}}
\def\Pen{P_{\rm en}}   
\def\Pnr{P_{\rm nr}}   
\def\Prel{P_{\rm rel}} 
\def\PrelOne{P_{\rm rel,lw}}
\def\PrelLw{P_{\rm rel,lw}}
\def\PrelTwo{P_{\rm rel,hi}}
\def\PrelHi{P_{\rm rel,hi}}
\def\ainj{a_{\rm inj}}
\def\amc{a_{\rm mc}}
\def\aint{a_{\rm int}}
\def\amax{a_{\rm max}}
\def\g{\gamma}
\def\GamGas{\g_{\rm g}}
\def\qTP{q_{\rm tp}}
\def\PRet{P_{\rm ret}}
\def\lGas{l_{\rm gas}}
\def\lDiff{l_{\rm D}}
\def\lmc{l_{\rm mc}}
\def\umc{u_{\rm mc}}
\def\MAZ{M_{\rm A0}}
\def\MAone{M_{\rm A1}}
\def\MSZ{M_{\rm S0}}
\def\MSone{M_{\rm S1}}
\def\Msub{M_{\rm sub}}
\def\Emax{E_{\rm max}}
\def\pmax{p_{\rm max}}
\def\Ptot{P_{\rm tot}}
\def\EnDen{\rho_{\rm E,tot}}
\def\EnSN{E_{\rm SN}}
\def\TISM{T_{\rm ISM}}
\def\downsound{c_{\rm S2}}
\def\tSed{t_{\rm Sed}}
\def\Ngone{N_{g1}}
\def\Ninj{N_{\rm inj}}
\def\pinj{p_{\rm inj}}
\def\pint{p_{\rm int}}
\def\PinjP{P^\prime_{\rm inj}}
\def\etaB{\eta_{\rm B}}
\def\lambdaB{\lambda_{\rm B}}
\def\L{{\lambda}}
\def\x#1{\!\times\!10^{#1}}
\def\Fesc{F_{\rm E}}
\def\FescR{F_{\rm E}(\Rtot)}
%

\begin{center}
%
{\LARGE \bf Photons and Particle Production in Cassiopeia A: 
Predictions from Nonlinear Diffusive Shock Acceleration}
\end{center}

\begin{center}
%
%
{\bf Donald C. Ellison$^{1}$, Philippe Goret$^{2}$, Matthew G. Baring$^{3}$}\\
{\bf Isabelle A. Grenier$^{2}$ and Pierre-Olivier Lagage$^{2}$}\\
{\it $^{1}$Physics Department, North Carolina State University,
Raleigh, NC 27695-8202, USA\\
$^{2}$Service d'Astrophysique, CEA, DAPNIA,
Centre d'Etudes de Saclay, 91191 Gif-sur-Yvette, France\\
$^{3}$Universities Space Research Association,
LHEA, NASA/GSFC, Greenbelt, MD 20771, USA}

\end{center}

\begin{center}
{\large \bf Abstract\\}
\end{center}
\vspace{-0.5ex}
We calculate particle spectra and continuum photon emission from the
Cassiopeia A supernova remnant (SNR). The particle spectra, ion and
electron, result from diffusive shock acceleration at the forward SNR
shock and are determined with a nonlinear \mc calculation. The
calculation self-consistently determines the shock structure under the
influence of ion pressure, and includes a simple parameterized
treatment of electron injection and acceleration. Our results are
compared to photon observations, concentrating on the connection
between the Radio and GeV--TeV gamma-ray range, and to cosmic ray ion
observations. We include new upper limits from the Cherenkov Array at
Th\'emis (CAT) imaging Cherenkov telescope and the Whipple 10m $\g$-ray
telescope at $> 400$ GeV.  These new limits support the suggestion
(\egc Cowsik \& Sarkar 1980; Allen \etal 1997) that energetic electrons
are emitting \syn\ radiation in an extremely high magnetic field ($\sim
1000 \muG$), far greater than values routinely assigned to the ISM, and
help to constrain our model. The large magnetic field allows
acceleration of cosmic ray ions to well above $10^{15}$ eV per nucleon
in the $\sim 300$ yr lifetime of Cas A.
%
\vspace{1ex}

\section{Introduction}
\label{intro.sec}
We describe particle acceleration in an expanding, spherical SNR blast
wave with a plane-wave, steady-state shock model. The justification for
using a steady-state calculation to model time-dependent SNRs is given
in \EB (1999) and the full details and assumptions of our model are
given in Baring \etal (1999). Briefly, the global SNR shock parameters
(\egc shock speed and radius as a function of remnant age) used as
input for the model, are estimated with a simple Sedov solution of the
evolving blast wave in an uniform medium.  Given the global shock
dynamics, we are able to calculate the shock structure and first-order
particle acceleration assuming that electrons and ions are accelerated
directly by the forward shock, leaving considerations of the reverse
shock, where substantial X-ray emission may originate, for later work.
Important limitations of our current model is that we do not consider
oblique magnetic field structures or include second-order Fermi
acceleration, which may be important in the low \alf Mach number shocks
($M_{A1} \sim 4$ for the parameters we show here) implied by the large
$B$-fields (see Bykov \& Uvarov 1999 for a model of electron injection
which does include 2nd-order acceleration).

For the particular case of Cas A, the forward shock may be interacting
with pre-supernova wind material swept into a relatively dense shell
(\egc Borkowski \etal 1996) which may account for the high magnetic
field if the stellar magnetic field was compressed along with the wind
material. Alternatively, the large field may be the result of
amplification by turbulent eddies (Keohane 1998 and references
therein).  If magnetic fields $B \sim 1000\muG$ dominate the
acceleration region, ions can be accelerated to well above $10^{15}$
eV/nuc. While the overall fluxes of these particles may not be
sufficient to provide the bulk of the cosmic rays at 1-10 GeV because
of the small numbers of particles swept up, particle spectra from young
SNRs might be harder than from older, slower SNRs and may dominate
above $\sim 10^{14}$ eV through the `knee' near $10^{15}$ eV.

Not only does the high $B$-field make it possible to produce cosmic
rays to $>10^{15}$ eV, but a homogeneous model with a single set of
parameters can be found which affords a reasonable fit to the {\it
intensities} of diffuse photon emission from radio to $\g$-rays.
Limits imposed by the radio and $\g$-ray observations allow us to place
constrains on the electron-to-proton ratios produced by shock
acceleration.  The detailed {\it shapes} of individual components (\egc
radio, X-ray), however, are difficult to model with a single set of
parameters.  Despite this limitation, important model and/or
environmental constraints can be inferred if it is assumed that the
relativistic electrons responsible for the radio synchrotron emission
also produce the diffuse infrared and X-ray continuum via the
\syn\ mechanism, (and in the same emitting volume). Likewise, if the
GeV and TeV $\gamma$-ray emission is dominated by \IC\ and
\brem\ emission (rather than \pion\ from energetic ions), it can be
assumed that these same relativistic electrons are responsible for the
GeV and TeV gamma rays. Assuming a homogeneous environment precludes,
of course, the modeling of emission from the lumpy morphology, knots,
ring structure, etc., that characterizes high spatial resolution
observations of Cas A.

\begin{figure}[t]
\centerline{\psfig{file=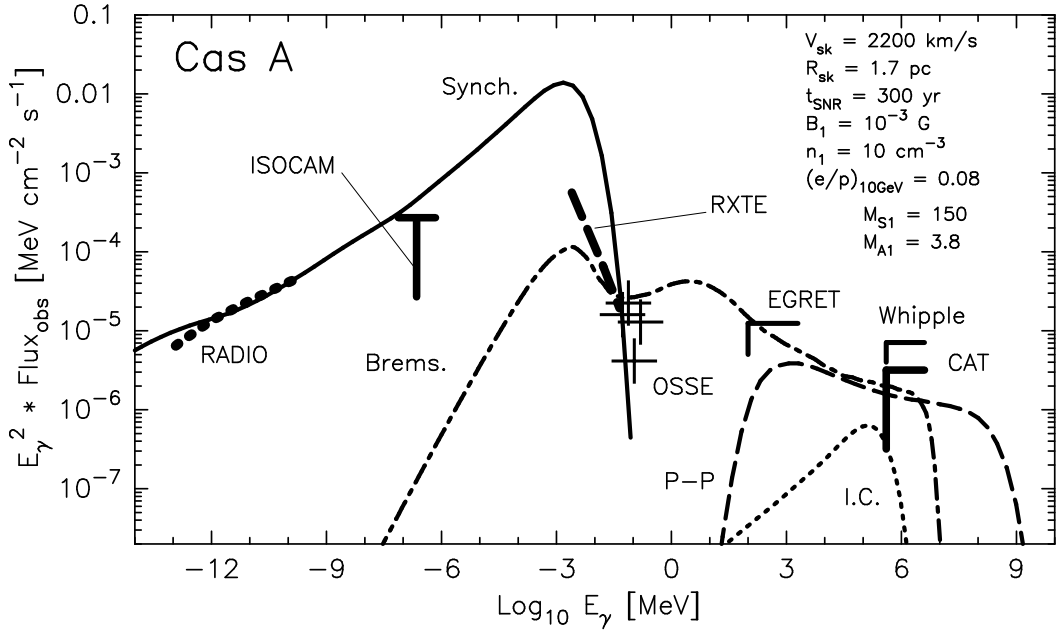,width=5.5truein}}
%
\caption{ {\it Cassiopeia A spectrum.
Observations are from Baars \etal (1977: radio); Lagage et al.  (1999,
in preparation: ISOCAM); The \etal (1996: RXTE and OSSE); Esposito
\etal (1996: EGRET); Lessard \etal (1999: Whipple); and Goret \etal
(1999: CAT).  The model photons come from a single set of proton,
helium, and electron spectra calculated with unshocked parameters shown
in the figure.  A single normalization has been applied to all
components to match the radio flux. The subscript `1' indicates
unshocked values and $M_{S1}$ and $M_{A1}$ are the sonic and \alf Mach
numbers. Note that the \brems\ emission cuts off at a much lower energy
than the \pion\ due to the \syn\ losses the electrons experience. In
these preliminary results, the \ic\ does not included \SSC\ although
this may dominate the \ic\ from the primordial background radiation.} }
\vspace{-0.4cm}
\end{figure}

\section{Results}
 \label{format.sec}
In Fig.~1 we compare our results to observations of Cas A from radio to
TeV $\g$-rays. We have a single normalization for all of the model
components and this has been chosen to match the radio flux.  Since the
same electron distribution that produces the radio emission can also
produce \brem\ and \IC\ emission at GeV-TeV energies, the combination
of the observed radio intensity and the low $\g$-ray upper limits
forces the conclusion that the magnetic field is exceedingly high.
Otherwise, the \brems\ and \IC\ emission at TeV energies would have
been observed.  This might not be the case if the radio emitting
electrons occupy a greater emission volume than GeV-TeV emitting
electrons, or that the lower energy (\iec radio-emitting) electrons can
preferentially sample regions of clumped magnetic field and/or
density.  Even if either of these cases arises, the \ic\ component,
which depends at most only weakly on the background particle density,
will set a lower limit on $B$ that remains well above standard ISM
values.  Note that the cosmic microwave background radiation forms the
seed for inverse Compton scattering in the present exposition; work is
in progress to include synchrotron self-Compton contributions, which
will tighten the constraints discussed here, implying even higher
$B$-fields.

The relative importance of the pion-decay emission depends sensitively
on the model parameters and we have not yet done a careful survey of
the parameter space.  For our preliminary results shown in Fig.~1, we
have chosen our electron injection parameters to give an electron to
proton ratio at relativistic energies, $\EoverP \simeq 0.08$,
somewhat above observed cosmic ray values.

\begin{figure}
\centerline{\psfig{file=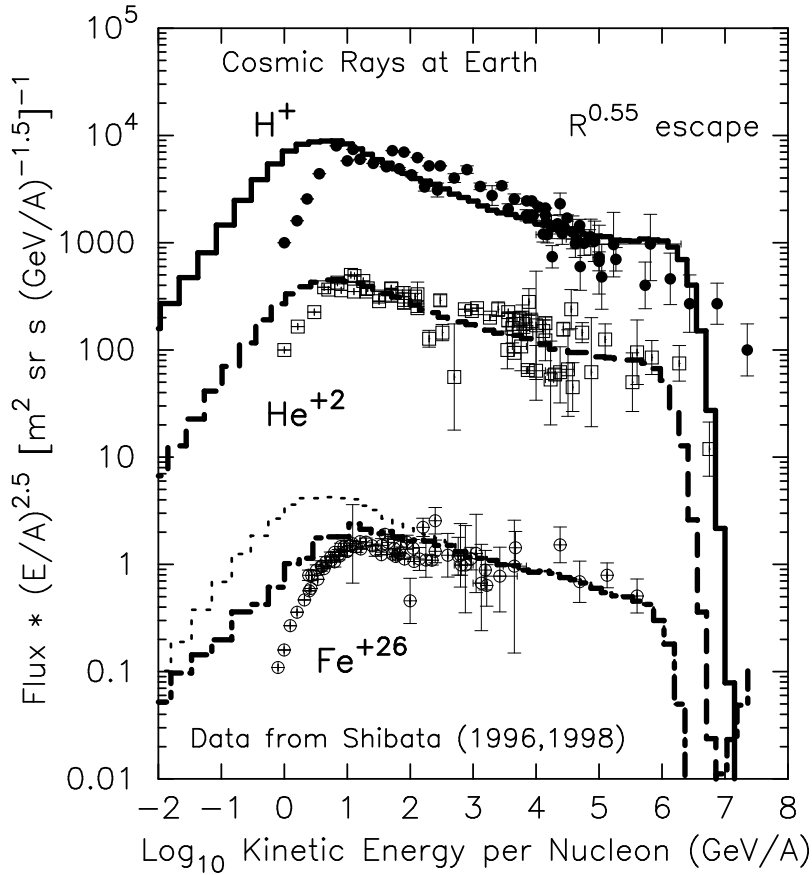,width=4.2in}}
%
\caption{{\it 
Cosmic ray spectra measured at the Earth (data from Shibata 1996, 1998)
compared to model spectra with same parameters as shown in Fig.~1. The
model spectra have been multiplied by Rigidity$^{-0.55}$ to account for
escape during propagation and have been normalized separately to match
the observations. The light dotted histogram is the model Fe spectrum
without nuclear destruction during propagation while the dot-dashed
histogram includes nuclear destruction. The turnover in the data below
$\sim 1$ GeV/A is from solar modulation, which is not included in the
model.}}
\vspace{-0.4cm}
\end{figure}
%
%

We again emphasize that even though our model is for plane-parallel,
steady-state shocks, detailed comparisons (Ellison \& Berezhko 1999)
with the spherically symmetric, time-dependent model of Berezhko
(1996; see also Berezhko, et al. 1996) show that the
steady-state and plane shock approximations do not seriously influence
the results as long as the diffusion length of the highest energy
particles is a small fraction of the shock radius, $\Rsk$, as should
be the case in Cas A and other young SNRs.  If Cas A is currently
interacting with a relatively dense shell of material formed from the
pre-SN wind (Borkowski \etal 1996), the Sedov solution can be replaced
by estimates of the ambient density, $n_1$, magnetic field, $B_1$,
$\Rsk$, and $\Vsk$, which translate to maximum particle energies.

\section{Conclusions}
\label{conclsion.sec}
The observed radio intensity of Cas A, combined with the EGRET and TeV
upper limits, imply magnetic fields $\gtrsim 1000\muG$. Fields this
large make it possible to accelerate cosmic ray ions to above $10^{15}$
eV/nuc in the $\sim 300\, {\rm yr}$ lifetime of the remnant, since the
time to shock accelerate ions of charge $Q$ to $\Emax$ is (\egc Baring
\etal 1999)
$$
\tacc \simeq
190 \,
\left ( \frac{\eta}{Q}  \right ) \,
\left ( \frac{\Vsk}{2000 \, {\rm km/s}} \right )^{-2} \,
\left ( \frac{B}{10^{-3} \, {\rm G}} \right )^{-1} \,
\left ( \frac{\Emax}{10^{15} \, {\rm eV}} \right )\,
{\rm yr}
\ .
$$
Here, $\eta$  is the number of gyroradii in a scattering mean free path
and is approximately one in the Bohm limit, which we assume in this
model.  In Fig.~2 we compare the model spectra, all multiplied by
$R^{-0.55}$ [$R=pc/(Qe)$] to model rigidity-dependent escape during
propagation, with cosmic ray observations.  No attempt has been
made to model the {\it abundances} of these components, each being
separately normalized to match the observations.  It's clear that the
spectra extend through the knee region.  Young SNR shocks sweep up far
less ISM material than older, larger shocks, but the high energy cosmic
rays produced may have flatter spectra than the bulk of the cosmic rays
accelerated by older SNRs which have weaker shocks and, if so, could
dominate the cosmic ray flux near the knee.  If this is the case,
spectral (and perhaps compositional) features should exist in the
cosmic ray spectra as these components become dominant.

If high $B$-fields are common in young SNRs, it has important
implications for $\g$-ray emissivity as well as cosmic ray production.
High fields imply that radio intensities will be high, concomitant with
relatively low relativistic electron fluxes (and presumably low
relativistic ion fluxes as well), lowering the $\g$-ray emissivity.
Therefore, radio loud SNRs may not be the best candidates for $\g$-ray
studies, and some other indicator may be required to guide
observational programs.

\hskip-36pt
%
\reff Allen, G.E. \etal 
\apjlet{97}{487}{L97}
\reff Baars, J.W.M., Genzel, R., Pauliny-Toth, I.I.K., \& Witzel,
A. \aa{77}{61}{99} 
\reff Baring, M. G., Ellison, D.~C., Reynolds, S.~P., Grenier, I.~A.,
 \& Goret P. \apj{99}{513}{311}
\reff Berezhko, E.G. \app{96}{5}{367}
\reff Berezhko, E.G., Yelshin, V.K. \& Ksenofontov, L.T. 1996, {\itt ZhETF},
{\bff 109}, 3.
%
%
\reff Borkowski, K.J., Szymkowiak, A.E., Blondin, J.M., \& Sarazin,
C.L. \apj{96}{466}{866}
\reff Bykov, A.M., \& Uvarov, Yu.A. \JETP{99}{88}{465}
\reff Cowsik, R., \& Sarkar, S. \mnras{80}{191}{855}
\reff Ellison, D.C., \& Berezhko, E.G. 1999, 26th ICRC (Salt Lake
City), OG 3.3.27.
%
%
\reff Esposito, J.A., Hunter, S.D., Kanbach, G., \& Sreekumar,
P. \apj{96}{461}{820} 
\reff Goret, P. \etal 1999, 26th ICRC (Salt Lake City), OG 2.2.18.
%
%
\reff Keohane, J.W. 1998, Ph.D. Thesis, University of Minnesota.
\reff Lessard, R. (Whipple Collaboration), 1999, Proc. 19th Texas Symposium, Paris 1998, in press.
\reff Shibata, T. 1996, {\itt Nuovo Cimento C}, {\bff 19}, 713.
\reff Shibata, T. 1998, private communication.
\reff The, L.-S., \etal \aasup{96}{120}{357} 
%
%

\end{document}